\documentclass{article}

\usepackage{lipsum}
\usepackage{lineno}
\usepackage{setspace}
\usepackage{natbib}
\usepackage{graphicx}
\usepackage{hyperref}
\usepackage{amsmath}
\usepackage[a4paper, total={6in, 8in}]{geometry}

\begin{document}
\begin{center}
{\LARGE \bfseries Modeling the Asymptotic Behavior of Higher-Order Aftershocks with Deep Learning\par\vspace{2\baselineskip}}

{\large Leila Mizrahi\textsuperscript{1}\textsuperscript{*} and Dario Jozinović\textsuperscript{1}\textsuperscript{*}}
\par\vspace{\baselineskip}
\textsuperscript{1} Swiss Seismological Service, ETH Zurich, Switzerland
\par \textsuperscript{*} Equal contribution. Corresponding author: leila.mizrahi@sed.ethz.ch
\end{center}

\begin{abstract}
Aftershocks of aftershocks - and their aftershock cascades - substantially contribute to the increased seismicity rate and the associated elevated seismic hazard after the occurrence of a large earthquake.
Current state-of-the-art earthquake forecasting models therefore describe earthquake occurrence using self-exciting point processes, where events can recursively trigger more events according to empirical laws.
To estimate earthquake probabilities within future time horizons of interest, a large number of possible realizations of a process are simulated, which is typically associated with long computation times that increase with the desired resolution of the forecast in space, time, or magnitude range.
We here propose a machine learning approach to estimate the temporal evolution of the rate of higher-order aftershocks.
For this, we train a deep neural network to predict the output of the simulation-based approach, given a parametric description of the rate of direct aftershocks.
A comparison of the two approaches reveals that they perform very similarly in describing synthetic datasets generated with the simulation-based approach.
Our method has two major benefits over the traditional approach. It is faster by several orders of magnitude, and it is not susceptible to being influenced by the presence or absence of individual `extreme' realizations of the process, and thus enables accurate earthquake forecasting in near-real-time. 

\end{abstract}

\section{Introduction}
    While deterministic earthquake prediction remains impossible, so-called earthquake forecasts can provide a probabilistic assessment of the possibility of future earthquakes, leveraging knowledge about the tendency of earthquakes to cluster in time and space.
    After the occurrence of a large earthquake, the seismicity rate of a region can be elevated by several orders of magnitude, and it may take years or even decades before it returns to what is commonly referred to as the background seismicity level.
    Fore-, main- and aftershocks can currently only be identified as such in hindsight, after an earthquake sequence has ended, and no differences in the physical mechanisms of these event types are presently known. 
    Thus, although the term ``aftershock'' is often perceived as describing an earthquake of a smaller magnitude than the large event that triggered it, there is always a chance that a large earthquake is followed by an even larger one.
    Additionally, smaller earthquakes may, in some cases, cause more damage than the largest, supposed main event, for example when they occur closer to populated cities or in environments where the local soil conditions amplify seismic waves more strongly.
    This highlights the importance of disseminating information about the possibility of future earthquakes, in particular during ongoing earthquake sequences.
    
    \cite{mizrahi2023unpublished} recently reviewed the current practices of such operational earthquake forecasting (OEF) systems in Italy, New Zealand, and the United States.
    In Italy, forecasts are provided regularly to civil protection; in New Zealand and in the US, earthquake forecasts are provided to the public and, in the case of the US, to the Bureau of Humanitarian Aid. 
    While the three agencies use different models to produce the information underlying their forecasts, all of them are or will in the near future be based at least partially on the Epidemic-Type Aftershock Sequence (ETAS; \citealp{ogata1988statistical}) model.
    The review of \cite{mizrahi2023unpublished} also elicited recommendations from an international group of experts on earthquake forecasting.
    The group agreed that the ETAS model would be their first choice when selecting ``a simple base model to produce earthquake forecasts that are useful for a maximum number of user groups''.

    \begin{figure}
        \centering
        \includegraphics[width=0.45\textwidth]{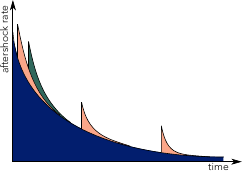}                
        \caption{Schematic illustration of the evolution of the aftershock rate, including higher-order aftershocks, with time.}
        \label{fig:h_o_aftershocks}
    \end{figure}
    
    A key attribute of the model is its capability to model higher-order aftershocks, that is, aftershocks of aftershocks and their aftershock cascades.
    This is in contrast to the \cite{reasenberg1989earthquake} model, which models directly triggered aftershocks of a main event using similar empirical laws as ETAS, but disregards the possibility that these aftershocks themselves may be large and trigger substantial amounts of aftershocks.
    \cite{helmstetter2003importance} theoretically analyzed the importance of modeling higher-order aftershocks for the correctness of seismicity rate forecasts, concluding that it is essential, especially for forecasting horizons that go beyond a few minutes.
    \cite{nandan2019forecastingrates} demonstrated the importance of modeling higher-order aftershocks by comparing the performance of different forecasting model types in California. 
    Accounting for higher-order aftershocks was also recognized by the expert group in the study of \cite{mizrahi2023unpublished} to be a crucial model ingredient for producing accurate forecasts.
    Because of the stochastic nature of ETAS-like models to describe seismicity, the spatiotemporal distribution of higher-order aftershocks is modeled in practice by generating a large number of possible realizations of the branching process. The resulting synthetic earthquake catalogs are used as a proxy to describe the distribution of possible outcomes.
    Depending on the desired resolution of the forecast and whether one is interested in a purely temporal or a spatiotemporal forecast, the number of generated catalogs can range from thousands \citep{harte2015log} to tens of thousands \citep{van2022prospective} to hundreds of thousands \citep{llenos2019ensembles, milner2020operational, nandan2021global, bayliss2022pseudo}.
    Producing these simulated event sets is typically associated with long computation times or computational resources required for parallelization.
    Approaches to approximate the behavior of higher-order aftershocks and hence avoid the computationally expensive simulation procedure have been proposed in the past (e.g., \citealp{helmstetter2003importance, saichev2006power, harte2017probability}).
    Yet, OEF forecasting systems still rely upon simulations to produce their forecasts \citep{mizrahi2023unpublished}.

    In this article, we propose a Machine Learning (ML) approach to model the asymptotic behavior of higher-order aftershocks, given an analytical description of the behavior of direct aftershocks.
    Machine Learning has been applied in different scientific fields to replace simulations, with the goal of reduced processing time \citep{kasim2021building, lam2023learning}. Examples in Seismology include speeding-up viscoelastic calculations \citep{devries2017enabling}, seismic wave simulation \citep{moseley2020deep}, synthetic earthquake waveform generation \citep{florez2022data}, etc. 
    In the field of earthquake forecasting, \cite{dascher2023using}, \cite{stockman2023forecasting}, and \cite{zlydenko2023neural} have recently proposed machine learning models as alternatives to ETAS models, achieving similar or slightly superior forecasting results.
    Unlike ETAS, these approaches offer the flexibility to incorporate additional datasets alongside earthquake catalogs, indicating potential for enhanced earthquake forecasting capabilities in the future.
    The goal of our study differs from that of \cite{dascher2023using}, \cite{stockman2023forecasting}, and \cite{zlydenko2023neural} in that we do not aim to fully replace ETAS models with ML models that are challenging to interpret and consequently difficult to communicate to non-specialized audiences.
    Our goal is to develop a deep neural network (NN) approach that emulates the results achieved through simulations of stochastic event sets, with a significant reduction of computational cost, while sticking to the simple and easily interpretable ETAS formulation of seismicity.
    We train the model on ETAS-simulated data and evaluate its output using commonly used metrics of both the ML and earthquake forecasting communities.

\section{Method and Data}
    \subsection{Objective}
        The ETAS model describes the seismicity rate at any given time and location as the sum of the background earthquake rate and the rate of aftershocks of all previous earthquakes \citep{ogata1988statistical, ogata1998space}.
        The aftershock rate is described using three components, which are usually assumed to be independent of each other: the aftershock productivity of an event depending on its magnitude, the temporal distribution of aftershocks, and their distribution in space.
        In this article, we will consider a temporal version of ETAS and disregard the spatial component for simplicity.
        If successful, our approach can be extended in future studies to similarly model the spatial behavior of higher-order aftershocks.
        The expected rate $\lambda(t)$ of direct aftershocks above a reference magnitude $m_{ref}$ of an event of magnitude $m$ is described as
        \begin{equation}
            \lambda(m, t) = k_0 \cdot e^{\alpha (m-m_{ref})} \cdot \frac{1}{(t + c)^p},
            \label{eq:1}
        \end{equation}
        where $t$ is the elapsed time in days since the origin time of the triggering earthquake.
        In ETAS simulations, the number of direct aftershocks and their occurrence times and magnitudes are sampled, whereupon their aftershocks are generated analogously, and this procedure is repeated until the number of simulated aftershocks becomes zero.
        Based on a large number of simulations, we obtain the expected rate of direct and higher-order aftershocks, which we denote by $\Tilde{\lambda}(m, t)$.
        Our objective is to model
        \begin{equation}
            \Tilde{\lambda}(m, t) - \lambda(m, t),
        \end{equation}
        given the parameters $k_0, \alpha, c, p$ and the magnitude $m$ of the initial earthquake which triggered the aftershock cascade.
        
    \subsection{Train and test data}
        We produce $\Tilde{\lambda} - \lambda$ data for 18,000 randomly selected parameter combinations. Table \ref{tab:params_used} shows the upper and lower bounds between which parameters are uniformly sampled. Note that $k_0$ and $c$ are sampled in log-space, and that the parameters describe the distribution of aftershocks above a reference magnitude of $m_{ref} = 2.2$. This choice of reference magnitude is arbitrary but does not affect the results in any other way than through the number of simulated events, which is similarly controlled by $k_0$.
        To avoid supercritical behavior in which each earthquake, on average, triggers more than one aftershock, leading to non-converging simulations, we only use parameter combinations for which the theoretical branching ratio $\eta$ is below 1.
        $\eta$ is the expected number of aftershocks triggered by an earthquake,
        \begin{equation}
            \eta = \int_0^{\tau} \int_{m_{ref}}^{\infty} \lambda(m, t) \cdot f_{GR}(m)  \,dm \,dt,
        \end{equation}
        where $f_{GR}(m)$ is the probability density function of earthquake magnitudes as described by the \cite{gutenberg1944frequency} law.
        For simplicity, we use a $b$-value of 1.
        To allow values of $p$ below 1, we use an upper limit $\tau$ of the temporal domain of the integral, which we set to 5 billion years.
        
        \begin{table}[]
            \centering
            \begin{tabular}{r|c|c}
                & Min & Max \\ \hline
                $\log_{10}(k_0)$ & -2.65 & -0.15 \\
                $\alpha$ & 0.5 & 2.4 \\
                $\log_{10}(c)$ & -4.5 & -1.0 \\
                $p$ & 0.8 & 1.2 \\ 
                $m$ & 4.5 & 6.5 \\
            \end{tabular}
            \caption{Parameter ranges used to produce ETAS simulations.}
            \label{tab:params_used}
        \end{table}

        For each parameter combination, we issue 20,000 simulations of aftershock cascades of a single initial event $e_0$ using the ETAS implementation of \cite{mizrahi2023etas}.
        From those simulations, we discard the simulated direct aftershocks of $e_0$ and use the remaining data to obtain an estimate of $\Tilde{\lambda} - \lambda$ by taking the mean number of events simulated within discretized time intervals.
        Each parameter combination then yields one row of data to be used by the ML model, where the parameters serve as model input, and the discretized $\Tilde{\lambda} - \lambda$ values are the expected model output.
        We consider 32 time intervals $[t_i, t_{i+1})$ of logarithmically increasing length, with $\log_{10}(t_i) = -4.0 + i\cdot 0.25$ days, with $i=0, \dots, 32$.
        The magnitude $m$ of $e_0$ has no other effect on the behavior of $\Tilde{\lambda} - \lambda$ than controlling the number of its direct aftershocks.
        Therefore, instead of providing $m$, we provide the logarithm of the expected total number of direct aftershocks of $e_0$ as a model input, $\log_{10}(n\textsubscript{DA})$.

    \subsection{NN Model}
        The resulting data are used for training the ML model. Explicitly, the inputs to the model are $\log_{10}(c), p-1, \alpha, \log_{10}(k_0),$ and $\log_{10}(n\textsubscript{DA})$, which are fed in as a NumPy array of shape (1,5). The outputs are the time-binned $\log_{10}(\Tilde{\lambda} - \lambda)$ values with 32 time bins. 
        Using the logarithm of the target quantity allows us to use the mean squared error (MSE) loss function that would otherwise overly optimize for higher values of $\Tilde{\lambda} - \lambda$. We split the data randomly into training (80\%), validation (10\%) and test (10\%) sets. All the results reported in the article are calculated on the test set unless differently noted. The validation set was used for model hyperparameter optimization.        
        
        The ML model we employ is a deep neural network with two hidden fully-connected layers with 64 and 40 neurons, respectively. Both hidden layers use the ELU activation function \citep{clevert2015fast}, and their weights are initialized using Glorot Uniform initialization \citep{glorot2010understanding}. We also tested using the more commonly used ReLU activation function \citep{nair2010rectified} but obtained better results using ELU. This can be explained by the dying ReLU problem \citep{lu2019dying}, which can significantly affect our network due to its small number of neurons. Dropout \citep{srivastava2014dropout} with a rate of 0.2 is applied to the output of the first hidden layer. The output of the second layer is then fed to another fully-connected layer with 32 neurons, which represents our model output. Our network has 4296 trainable parameters in total. The model has been trained using Keras \citep{chollet2015keras} available through TensorFlow \citep{tensorflow2015-whitepaper}.

        We use a batch size of 32 for training and train the model until the validation loss does not improve for 30 epochs (with a maximum of 500 epochs in total). When the training is stopped, the model with the lowest validation loss is restored and saved. The optimizer chosen is Adam with an initial learning rate of $10^{-2}$, which is reduced during training by multiplying the current learning rate with a factor of 0.2 whenever the validation loss does not improve for 15 epochs (with $10^{-7}$ being the lowest possible learning rate). We use the mean squared error (MSE) as the loss function.

        The validation loss stops improving after 202 epochs, which takes 82 seconds on a personal laptop (Thinkpad T14 Gen 2a) CPU. The training set MSE was 0.0084, as was the validation set MSE.

        In addition to the NN, we also tried to use a multi-output linear regression to model the time-binned $\log_{10}(\Tilde{\lambda} - \lambda)$ values, which would yield a more easily interpretable model with an analytic form. However, we found that the resulting MSE was an order of magnitude higher than that of the NN, meaning that the non-linearity modelling capability of deep neural networks provides a substantial improvement to the description of the temporal evolution of higher-order aftershocks.
        
    \subsection{Model evaluation}
        Besides evaluating the model based on the MSE achieved on the test data, we also compare the ML model's performance to the simulation-based approach using the information gain (IG) metric, which is commonly applied to compare the skills of different forecasting models \citep{harte2005entropy, rhoades2011efficient, zechar2010collaboratory, strader2017prospective}.
        We conduct synthetic tests as follows.
        For parameter combinations that were part of the test data described in the previous section, and hence were not used to train the ML model, we produce 100 simulations of aftershock cascades. For each individual simulation, direct aftershocks are removed and the remaining higher-order aftershocks are counted within the same time bins that were used to prepare the training data.
        Each simulation represents a possible realization of the known underlying branching process.
        We then quantify how well the expected (mean) number of higher-order aftershocks describes the individual realizations, both when the expected value is calculated through 20,000 ETAS simulations, and when it is calculated with the ML model.
        For a specific realization $r$, the information gain per earthquake (IGPE) of the ML approach compared to the simulation-based approach is given as

        \begin{equation}
            \text{IGPE}_r = \sum_{i=0}^{31}\frac{1}{n_{r}^i + 1}\left(\ln\left(P_{ML}^i(n_{r}^i | {\Theta_r})\right) - \ln\left(P_{sim}^i(n_{r}^i | {\Theta_r})\right)\right),
            \label{eq:IG}
        \end{equation}
        where $n_{r}^i$ is the number of higher-order aftershocks in realization $r$ in the $i^{th}$ time interval $[t_i, t_{i+1})$,  and $P_{ML}^i(n | \Theta_r)$ and $P_{sim}^i(n | \Theta_r)$ are the probability mass functions (PMF) of observing $n$ events in the $i^{th}$ time interval based on the expected values calculated by the ML-based and simulation-based models, respectively, with $\Theta_r$ representing the parameters used to generate realization $r$.
        Because large absolute information gain (IG) values mostly arise when `observed' event numbers are large, we consider the IGPE by dividing the IG by $n_{r}^i + 1$, where the $+1$ enables the inclusion of instances with no events in our comparison.
        To obtain the PMFs from modeled expected event numbers, we assume a Poisson distribution.
        This is a simplification that is known not to be ideal for modeling earthquake number distributions.
        \cite{nandan2019forecastingfull} showed that forecasting the full distribution of the number of future earthquakes better captures the large uncertainty associated with it than solely forecasting its mean.
        However, the Poisson assumption is still widely used in forecast model test metrics, for instance, in those proposed by the Collaboratory of the Study of Earthquake Predictability (CSEP, \citealp{savran2022pycsep}).
        For the purpose of this article, we argue that using the Poisson assumption enables a first verification of the suitability of our proposed approach to model higher-order aftershock rates. 
        If the approach proves effective, it can be extended to model uncertainties, too.

\section{Results and Discussion}
    \subsection{NN model performance}
    
    \begin{figure*}[ht]
        \centering
        \includegraphics[width=\textwidth]{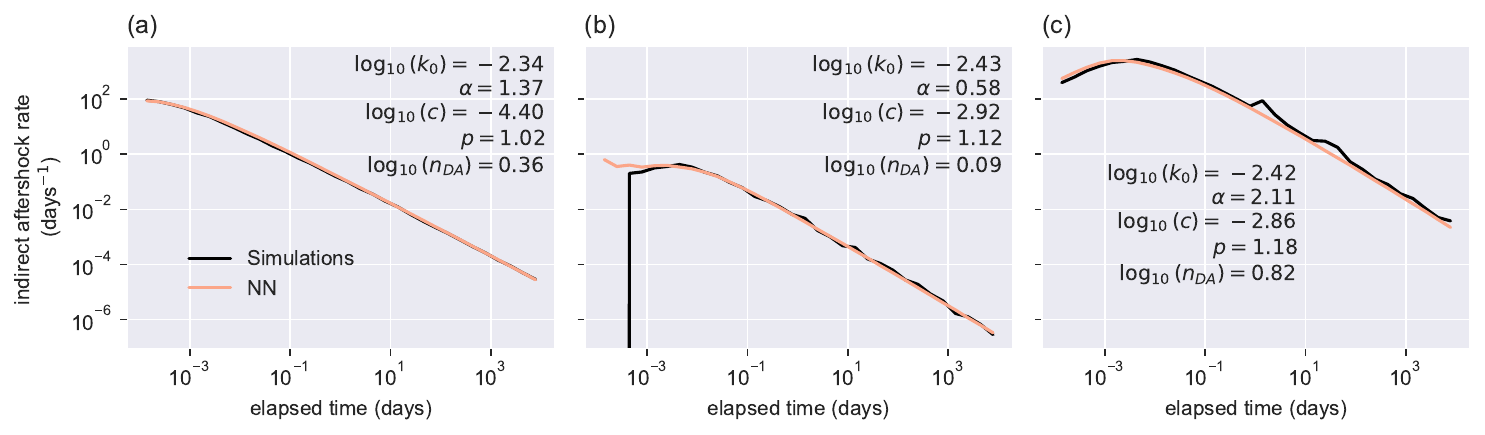}
        \caption{Examples of the temporal evolution of expected higher-order aftershock rates for different parameter combinations, as modeled with the simulation-based and NN-based approaches. The examples shown are those where the MSE ranks at the 5$^{th}$ (a), 50$^{th}$ (b), and 95$^{th}$ (c) percentiles among the test data.}
        \label{fig:sim_vs_nn}
    \end{figure*}

    When applied to the test set, the model achieves a MSE of 0.0093. The MSE of the test set is slightly higher than the MSE of the training and validation sets. This suggests that there is little overfitting of the model to the training data, making us confident in the generalization abilities of the model.
    
    The mean time needed to estimate the discretized $\Tilde{\lambda} - \lambda$ values for one set of parameters is $813\  \mu s$. When applied to a list of parameter sets, the time does not increase substantially (e.g., for 18,000 parameter combinations, the run time is $5.09\ ms$). 
    
    The run time of the classical approach using simulations heavily depends on the number of simulations done, the lower magnitude threshold of simulated events, the number of events of which aftershock cascades are being simulated, etc. 
    The code used to create the training data for this study was not optimized for speed. Simulations were done in space-time, yet only temporal information was used. 
    The lower limit of simulated magnitudes was not optimized to minimize computation time without a loss of forecast precision.
    Some parameter combinations may potentially produce unrealistically many aftershocks and lead to long computation times that would not be expected in a realistic situation. On the other hand, the ML model was run on a personal laptop, while the simulations were run on the Euler cluster of ETH Zürich, making the run times between the two approaches even less comparable.
    Nevertheless, an assessment of the computation time required to generate the training data versus the time needed by the ML model can provide an idea of the scale at which the two approaches differ in terms of computation time.
    The mean run time of the classical approach with 20,000 simulations was 54 minutes to simulate over half a million events on average, out of which 30\% were higher-order aftershocks.
    This means that the usage of the ML model could speed up single calculations by several orders of magnitude, with the mean speedup of $4 \cdot 10^{6}$ times in the case of the calculations done for the present analysis. 
    This is a significant improvement, useful for researchers with low computing power available, and for faster, near-real-time, higher-order aftershock forecasting.

    The performance achieved on the training, validation and test datasets suggests that the model performs well at estimating $\Tilde{\lambda} - \lambda$.
    Figure \ref{fig:sim_vs_nn} shows the comparison of the expected higher-order aftershock rate obtained with the two approaches for three examples of parameter combinations.
    The examples shown are those with a MSE ranking at the 5$^{th}$, 50$^{th}$, and 95$^{th}$ percentiles among the test data.
    While the lines in Figure \ref{fig:sim_vs_nn}(a) overlap almost completely,  
    a distinct contrast between simulation-based and NN-based expected rates is evident in Figure \ref{fig:sim_vs_nn}(b), where the simulation-based method yields a rate of zero for very early time intervals, while the NN-based approach does not.
    Because the initial time intervals are shorter than the later ones, parameter combinations with relatively small event numbers may lead to none of the simulations containing any events in these early time intervals.
    In such a case, a large number of realizations is required to obtain a robust estimate of the average event rate with the simulation-based approach.
    Moving to Figure \ref{fig:sim_vs_nn}(c), the two curves appear similar, but the NN-based curve is smoother than the simulation-based one.
    The spikes obtained with the simulation-based approach are a consequence of (possibly several) individual simulations with particularly large events triggering their own, highly productive aftershock cascades.
    These observations reveal an additional potential benefit of the NN-based approach: it captures the collective behavior of higher-order aftershocks derived from a variety of parameter combinations.
    In contrast to the simulation-based approach, it is therefore not prone to be influenced by individual realizations of the branching process, and can provide small but nonzero expected event numbers when the simulation-based approach would require an exceedingly large number of simulations.

    \subsection{Model dependency on input parameters}
    \begin{figure*}[ht]
        \centering
        \includegraphics[width=\textwidth]{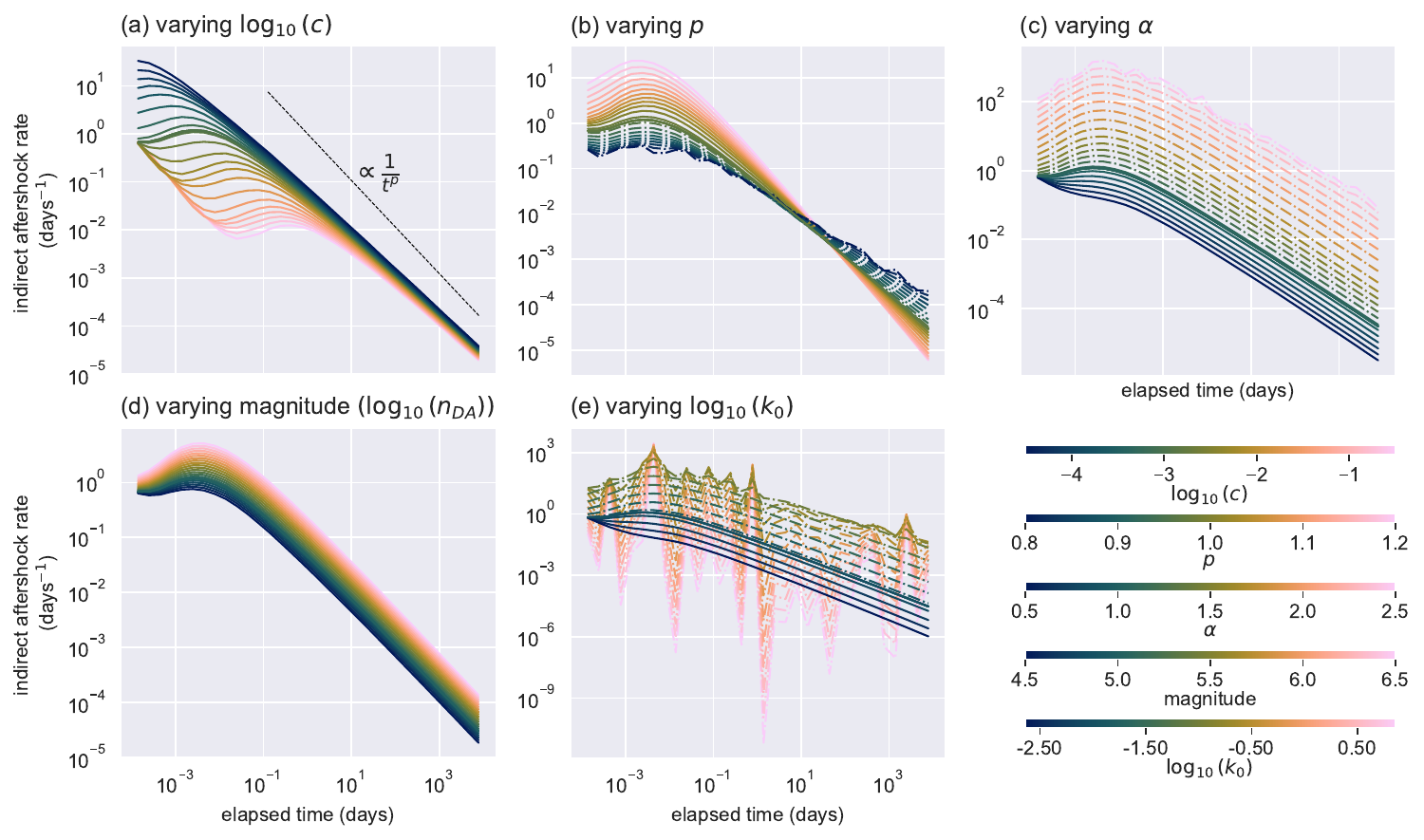}
        \caption{Dependence on input parameters of the temporal evolution of higher-order aftershock rate, as modeled by the NN approach. Starting from a reference set of parameters ($\log_{10}(c) = -2.95, p=0.97, \alpha = 1.07, \log_{10}(k_0) = -2.01, m = 5.0$), panels (a)-(e) show how $\Tilde{\lambda}-\lambda$ varies when $\log_{10}(c), p, \alpha, m,$ and $\log_{10}(k_0)$ vary, respectively. Dash-dotted lines indicate that the respective parameter combinations yield branching ratios of $\eta > 1$ and thus are outside of the parameter space covered by the training data. The dotted line in (a) represents the temporal decay of the first-order aftershock rate prescribed through the parameter $p$.}
        \label{fig:par_var}
    \end{figure*}

    An often acknowledged disadvantage of machine learning models is their so-called `black-box' nature.
    While Equation (\ref{eq:1}) describes a long-established empirical relationship with only four parameters, our ML model has 4296 parameters that were optimized during its training, making it hard for a human mind to comprehend how the NN derives higher-order aftershock rates over time from the five input parameters.
    To nevertheless explore the dependence of $\Tilde{\lambda} - \lambda$ on the NN's input parameters, we show in Figure \ref{fig:par_var} how the model output varies when one input parameter changes while the others are fixed.
    Starting from a reference set of parameters that correspond to those inferred for California by \cite{mizrahi2021effect}, panels (a-e) of Figure \ref{fig:par_var} show $\Tilde{\lambda}(t) - \lambda(t)$ with varying $\log_{10}(c), p, \alpha,$ magnitude $m$ of the initial event and with it its number of expected direct aftershocks $n_{DA}$, and $\log_{10}(k_0)$, respectively.
    Qualitatively, the behavior of the NN model is reasonable.
    \begin{itemize}
        \item (a): Larger values of $c$ lead to a delayed peak of the higher-order aftershock (HOA) rate, caused by a delayed start of the decay of the direct aftershock (DA) rate.
        \item (a): The overall HOA rate decay is slightly slower than that of the DA rate, as illustrated by the dotted line, which represents the DA rate decay.
        \item (b): A slower DA rate decay (smaller $p$) results in a slower HOA rate decay.
        \item (c): Because any earthquake can trigger aftershocks of any magnitude, the magnitude dependency of the aftershock productivity described by the parameter $\alpha$ only manifests in larger average aftershock numbers and the associated cascading effects thereof.
        An increase in $\alpha$ thus leads to an overall increase in event rate and to an exaggeration of the shape of the HOA rate evolution curve.
        \item (d-e): The impact of the overall expected number of DA, governed by the magnitude $m$ of the initial event (d), and the influence of the number of aftershocks unrelated to the magnitude of the triggering event, governed by $k_0$ (e), on the HOA rate are, as expected, similar to that of $\alpha$.
        \item Dash-dotted lines in (b), (c), and (e) indicate parameter combinations with branching ratios above 1; hence, these are outside the parameter space used to train the model.
        It appears that the model yields somewhat plausible results when extrapolating in the vicinity of the training parameter space, but with increasingly unrealistic inputs, the model outputs become increasingly unstable.
        This is not a flaw of the model but merely a reminder not to blindly apply it with input parameters that are not covered by the training data.
        Neural networks often have problems extrapolating on data far away from their training distribution, and this behavior is therefore expected.
        For instance, the dash-dotted lines in (e) reflect parameter combinations with branching ratios of up to over 600, where the shape of the HOA rate curve starts exhibiting erratic behavior above a branching ratio of around 8.0.
    \end{itemize}

    \subsection{Synthetic test results}

\begin{figure}[ht]
        \centering
        \includegraphics[width=0.7\textwidth]{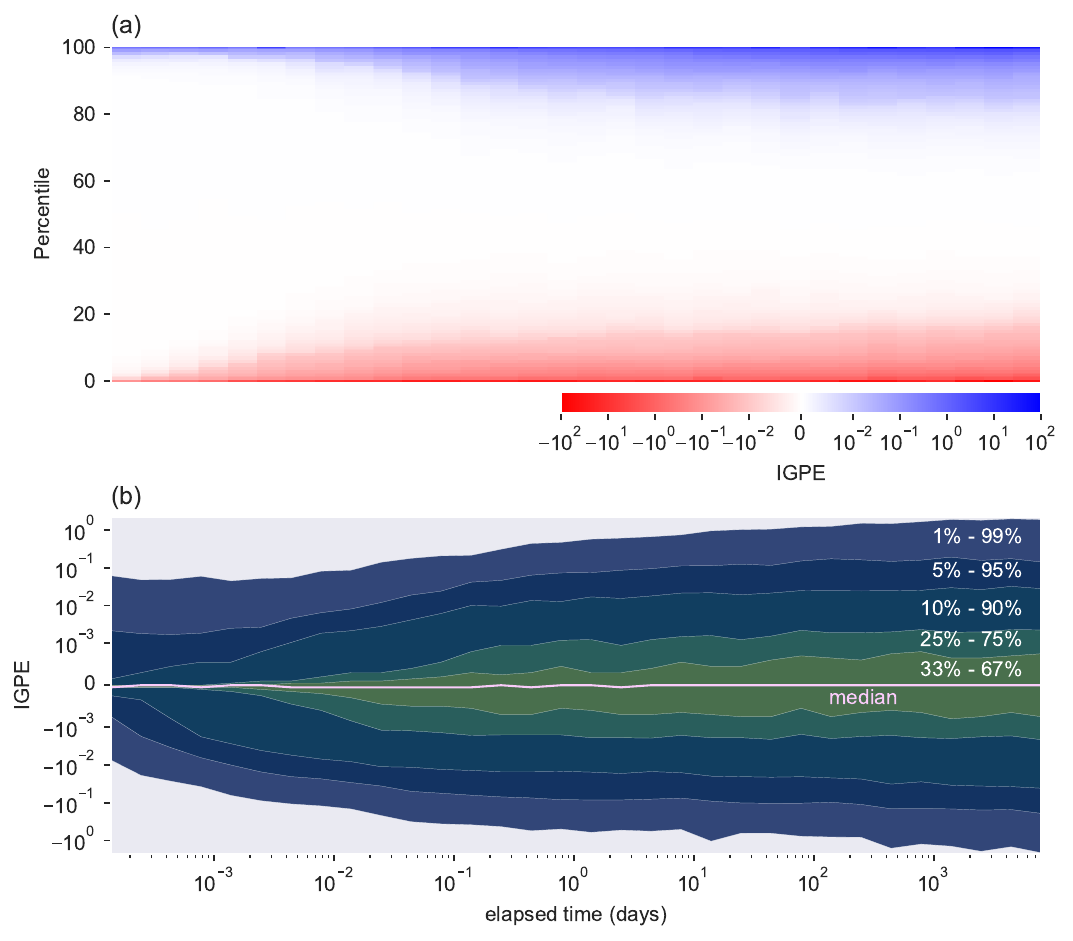}
        \caption{
        Information gain per earthquake (IGPE) values depending on the elapsed time since the initial event $e_0$.
        (a) 0$^{th}$ to 100$^{th}$ percentile in steps of 1\%, per time bin $t_i$.
        (b) Confidence intervals for IGPE versus time.
        }
        \label{fig:igpe_in_time}
    \end{figure}

    Figure \ref{fig:igpe_in_time} shows the performance difference between the NN model and the simulation-based approach in the synthetic test, depending on the time interval $t_i$ for which expected HOA rates are calculated.
    Thus, the information gain per earthquake (IGPE) is analyzed before it is summed over time bins as described in Equation (\ref{eq:IG}).

    The main result derived from Figure \ref{fig:igpe_in_time} is that the two models, NN-based and simulation-based, perform extremely similarly.
    The vast majority (80\%) of IGPE values have absolute values of less than 0.01, and over 53\% of absolute values are below $10^{-3}$.
    This means that the NN-based model performs as well as the traditional simulation-based approach at estimating expected HOA rates.
    Although it is unexpected that the NN-based model can outperform the simulation-based model, the output of which it was trained on, there are some instances in which the NN exhibits large IGPE values, and thus performs better than its benchmark. 
    On the other hand, the traditional approach is more successful in other instances, leading to negative IGPE values.
    Figure \ref{fig:igpe_in_time}(a) shows the $0^{th}$ to $100^{th}$ percentile IGPE values in each of the 32 time intervals after the occurrence of the initial event $e_0$, and Figure in \ref{fig:igpe_in_time}(b), IGPE confidence intervals are shown per time interval.
    In the early time intervals, the confidence intervals are very narrow, and they become larger for later time intervals.
    This is most likely due to the increasing uncertainty in the evolution of an earthquake sequence with time.
    On one hand, this means that the training data provided to the NN becomes more volatile and thus more difficult to learn. At the same time, this increased volatility indicates that the benchmark model itself becomes less stable as the forecasting horizon extends further to the future. 
    The test data, produced by simulations, suffers from the same instability.
    If the NN model is indeed able to learn a smoothed version of the simulation-based output, it would still be expected that, by chance, the simulation-based approach sometimes happens to better describe an `observation'.
    Large (positive and negative) IGPE values can be interpreted to arise from `exceptional' test datasets, where sometimes such exceptional cases may coincidentally be covered by the simulation-based approach, leading to negative IGPE, while in others, the simulation-based approach may not capture the exceptional case. Instead, the smoothed NN-based HOA rate better describes it, resulting in positive IGPE values.
    Therefore, if the NN model does more accurately capture the asymptotic behavior of HOA than the simulation-based approach, positive and negative IGPE values will necessarily be obtained in experiments such as the one presented here.

    \begin{figure}[ht]
        \centering
        \includegraphics[width=0.7\textwidth]{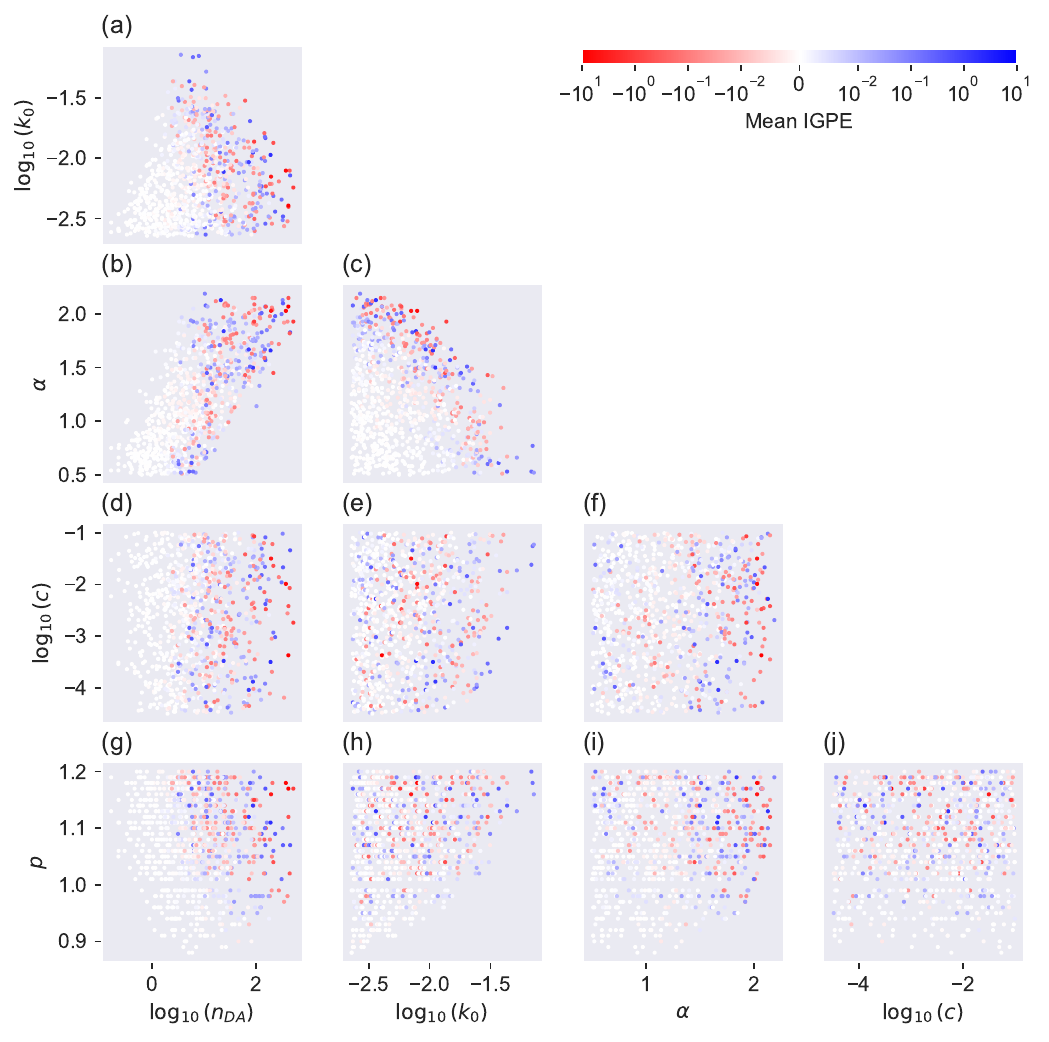}
        \caption{
        Scatter plot matrix of mean IGPE values per parameter combination, for all pairs of parameters among the inputs to the NN model, $\log_{10}(n\textsubscript{DA}), \log_{10}(k_0), \alpha, \log_{10}(c), p$.
        }
        \label{fig:migpe_scatter}
    \end{figure}

    To rule out that highly positive and negative IGPE values are systematically achieved in certain input parameter ranges, Figure \ref{fig:migpe_scatter} shows scatter plots of mean IGPE values per input parameter combination, for all possible pairs of parameters among the inputs to the NN model.
    For this, IGPE values are calculated for individual realizations as described in Equation (\ref{eq:IG}), and for each parameter combination, the mean over 100 realizations is taken to produce one dot in the scatter plots.

    Several interesting observations can be made based on Figure \ref{fig:migpe_scatter}.
    There is a clear trend showing in (a, b, d, g) that a small number of expected direct aftershocks leads to IGPE values close to zero, and larger $\log_{10}(n_{DA})$ values produce larger IGPE values, without there being a trend in their sign.
    This is in line with our interpretation given before that larger volatility in the test data, which results from larger overall event numbers, explains larger absolute IGPE values.
    Similarly, a combination of large $k_0$ and large $\alpha$ produces more volatile test data and thus shows larger absolute IGPE values in (c).
    For the other parameters, no obvious systematic behavior can be inferred, but Figure \ref{fig:migpe_scatter} offers insight into the input parameter range encompassed by the study.
    Although all parameters were sampled uniformly, clear patterns arise as to which parameter combinations are deemed unnatural (i.e., yield branching ratios $>1$ ) and, consequently, left unexplored.

\section{Conclusion}
    We propose a machine learning-based approach to model the asymptotic temporal behavior of higher-order aftershocks, given a parametric description of the behavior of direct aftershocks.
    This task is highly relevant in operational earthquake forecasting because higher-order aftershocks can contribute substantially to the temporally elevated seismic hazard following large events.
    The main advantage of our proposed approach over the traditionally used approach, which relies on thousands of simulations of possible evolutions of an earthquake sequence, is its computational efficiency.
    Using our model leads to a computation speedup of several orders of magnitude, enabling near-real-time earthquake forecasting, which is particularly useful early on in an unfolding earthquake sequence.
    We evaluate the model using different metrics that are commonly used in the machine learning and earthquake forecasting communities:
    The ML model achieves a low mean squared error on the test dataset that was not used to train the model, and an information gain per earthquake of virtually zero when it is compared to the traditional approach in synthetic tests.
    These results suggest that the ML model performs as well at forecasting higher-order aftershock numbers as the simulation-based approach, at a significantly increased computational efficiency.
    The very similar performance of the two approaches is observed independently of the temporal distribution of direct aftershocks.
    
    A more detailed investigation of the model's performance reveals that our proposed approach possibly more accurately describes the asymptotic rate of higher-order aftershocks than the simulation-based approach does, because it is not sensitive to individual `extreme' realizations of the assumed underlying branching process or a lack thereof.

    These findings open a new and exciting avenue for earthquake forecasting research. For practical applications, our proposed machine learning approach should be extended to model not only the mean but the full distribution of higher-order aftershock numbers, to avoid the Poisson assumption. Furthermore, in addition to the temporal aspect of higher-order aftershock occurrence investigated in the present work, their occurrence in space can be treated analogously. In addition to the improvement of aftershock forecasting capabilities of the machine learning model, further effort should be invested into its interpretability. For instance, simpler and thus more interpretable ML models or methods that help with the interpretation of black-box models, e.g., LIME \citep{ribeiro2016should} or SHAP \citep{lundberg2017unified} approaches, can be explored.

\section*{Data and Resources}
The code and the trained model are available on GitHub \href{https://github.com/djozinovi/cascadeNN}{cascadeNN} and  \href{https://doi.org/10.5281/zenodo.10304408}{Zenodo}. The data used in this paper (split into the training, validation, and test set) are available through \href{https://doi.org/10.5281/zenodo.10304408}{Zenodo}.

\section*{Declaration of Competing Interests}
The authors declare no competing interests.

\section*{Acknowledgements}

This work was supported by the EU project “A Digital Twin for Geophysical Extremes” (DT-GEO) (No: 101058129) and by the Swiss Federal Nuclear Safety Inspectorate (ENSI).

\bibliography{mybibfile}

\end{document}